\documentclass[10pt,aps,prb,twocolumn,preprintnumbers,amsmath,amssymb,floatfix,showpacs]{revtex4-1}
\usepackage[latin1]{inputenc}
\usepackage[T1]{fontenc}
\usepackage[english]{babel}
\usepackage[pdftex]{graphicx}
\usepackage{amsmath}
\usepackage{times}
\usepackage[usenames,dvipsnames,svgnames,table]{xcolor}
\usepackage[colorlinks,plainpages=false,linkcolor=blue,urlcolor=blue,citecolor=blue,pdfpagemode=UseNone]{hyperref}

\renewcommand{\AA}{\text{\r{A}}}

\newcommand\Vek[1]{\vec{#1}}

\begin{document}

\title
{
\boldmath
Confinement- and strain-induced enhancement of thermoelectric properties in LaNiO$_3$/LaAlO$_3(001)$ superlattices
}

\author{Benjamin Geisler}
\email{benjamin.geisler@uni-due.de}
\affiliation{Fakult\"at f\"ur Physik, Universit\"at Duisburg-Essen and Center for Nanointegration (CENIDE), Campus Duisburg, Lotharstr.~1, 47048 Duisburg, Germany}
\author{Rossitza Pentcheva}
\email{rossitza.pentcheva@uni-due.de}
\affiliation{Fakult\"at f\"ur Physik, Universit\"at Duisburg-Essen and Center for Nanointegration (CENIDE), Campus Duisburg, Lotharstr.~1, 47048 Duisburg, Germany}

\date{\today}

\begin{abstract}
By combining \textit{ab initio} simulations including an on-site Coulomb repulsion term
and Boltzmann theory, we explore the thermoelectric properties of (LaNiO$_3$)$_n$/(LaAlO$_3$)$_n$(001) superlattices ($n=1,3$)
and identify a strong dependence on confinement, spacer thickness, and epitaxial strain.
While the system with $n=3$ shows modest values of the Seebeck coefficient and power factor,
the simultaneous reduction of the LaNiO$_3$ region and the LaAlO$_3$ spacer thickness to single layers
results in a strong enhancement, in particular of the in-plane values.
This effect can be further tuned by using epitaxial strain as control parameter:
Under tensile strain corresponding to the lateral lattice constant of SrTiO$_3$
we predict in- and cross-plane Seebeck coefficients of $\pm 600$~$\mu$V/K and an in-plane power factor of $11$~$\mu$W/K$^2$cm
for an estimated relaxation time of $\tau = 4$~fs around room temperature.
These values are comparable to some of the best performing oxide systems
such as La-doped SrTiO$_3$ or layered cobaltates
and are associated with the opening of a small gap ($0.29$~eV)
induced by the concomitant effect of octahedral tilting and Ni-site disproportionation.
This establishes oxide superlattices at the verge of a metal-to-insulator transition driven by confinement and strain
as promising candidates for thermoelectric materials.
\end{abstract}


\maketitle

\section{Introduction}

Transition metal oxides are an important class of materials  for thermoelectric applications
due to their chemical and thermal stability
and environmental friendliness, but also to the prominent role of electronic correlations.~\cite{HeLiuFunahashi:11, HebertMaignan:10}
Considerable experimental and computational~\cite{Gorai:17} research aims at finding oxide thermoelectrics
with improved performance, mostly among bulk materials~\cite{Ong:10, Klie:12}
by doping~\cite{Xing:16,Garrity:16,LamontagneGrunerPentcheva:16,Okuda:01} or strain.~\cite{Gruner:15}
Nanostructuring and dimensional confinement are alternative strategies
to  enhance the thermoelectric properties.~\cite{SizeEffectTE:16, HicksDresselhaus:93}
To this end, modern layer-by-layer growth techniques
allow for an atomic-scale design of artificial transition metal oxide heterostructures
that display exotic characteristics,
notably different from their bulk components.~\cite{Hwang:12, Mannhart:10, Chakhalian:12, RENickelateReview:16, OxideRoadmap:16}
Concerning the thermoelectric response, a particularly promising system is $\delta$-doped SrTiO$_3$ (STO).~\cite{Ohta:07, Stemmer:10}
On the theoretical side, the role of confinement has been addressed in a few cases,
e.g., in LaAlO$_3$/SrTiO$_3(001)$ and $\delta$-doped STO superlattices (SLs).~\cite{Pallecchi:15, Filippetti:12, Delugas:13, GhosezSTO:16} Recently, the design of $n$- and $p$-type oxide thermoelectrics, exploiting the selective polarity of their interfaces, was proposed in (LaNiO$_3$)$_3$/(STO)$_3(001)$ SLs.~\cite{Geisler-LNOSTO:17}

Another system that has been in the focus of intensive research comprises the  paramagnetic correlated metal LaNiO$_3$ (LNO)~\cite{LNO-STO-FermiSurfaces:09, Ouellette:10, Sreedhar:92, Zhou:03} and the band insulator LaAlO$_3$ (LAO). In 2008  Chaloupka and Khaliullin~\cite{ChaloupkaKhaliullin:08} proposed that a cuprate-like electronic structure~\cite{Hansmann:09} can be realized in (LaNiO$_3$)$_1$/(La$B$O$_3$)$_1(001)$ SLs
($B$ being a trivalent cation such as Al) by using confinement and strain. Subsequent studies have shown that a single or a double LNO layer confined in LAO undergoes a metal-to-insulator transition (MIT) for tensile strain~\cite{ABR:11, Freeland:11, LiuChakhalian:11} and exhibits magnetic order.~\cite{Frano:13, Boris:11,Puggioni:12,LuBenckiser:16}
Additionally, considerable orbital polarization can be obtained through strain~\cite{Frano:13} or chemical control by using different $B$ counterions (B, Al, Ga, In).~\cite{Han-ChemCtrl:10}
With increasing LNO thickness the metallic behavior common to bulk LNO is restored, e.g., in (LNO)$_n$/(LAO)$_3(001)$ SLs on STO ($n=3,5,10$)~\cite{ABR:11, LiuChakhalian:11} or (LNO)$_4$/(LAO)$_4(001)$ SLs on STO.~\cite{Benckiser:11, LNOLAO-4-4-ParkMillisMarianetti:16}
In contrast to
LNO/STO$(001)$ SLs,~\cite{Geisler-LNOSTO:17}
in LNO/LAO$(001)$ SLs there is no polar discontinuity at the interface,
and unlike LNO/LaTiO$_3(001)$ SLs,~\cite{ChenMillisMarianetti:13}
no charge transfer across the interface arises,
leaving the nickelate in LNO/LAO$(001)$ SLs undoped.~\cite{ChargeTransferChenMillis:17}

Despite this rich electronic and magnetic behavior of nickelate SLs,
the thermoelectric properties have not been addressed so far.
Here we focus in particular on the effect of confinement and epitaxial strain
in (LNO)$_n$/(LAO)$_n(001)$ SLs with $n=1, 3$.
We provide a detailed analysis of the in- and cross-plane electronic transport
by using Boltzmann theory in the constant relaxation time approximation
and determine the Seebeck coefficients and estimates for the attainable power factor,
focusing on around-room-temperature applications.
We find that the degree of quantum confinement in the vertical direction
strongly affects the thermoelectric response,
particularly in-plane.
We furthermore disentangle the role of octahedral tilting in the previously predicted Ni-site disproportionation
under tensile strain.~\cite{ABR:11}
We show that the sensitivity of the MIT
and of the band velocities near the emerging band gap
to epitaxial strain
can be used to optimize the thermoelectric performance of such SLs.
The room-temperature values for the Seebeck coefficient and the power factor are compared to other topical oxide systems
like La-doped STO~\cite{GhosezSTO:16, Okuda:01, Stemmer:10}
or layered cobaltates.~\cite{YordanovCCO:17, GhosezCCO:17, CCO-Schwingen-APL:17}
By contrasting the ultrathin SLs
to the metallic longer-period (LNO)$_3$/(LAO)$_3(001)$ case
we point out the relevance of the MIT
in obtaining a high thermoelectric response.

\section{Methodology}

We have performed first-principles calculations in the framework
of spin-polarized density functional theory~\cite{KoSh65} (DFT)
as implemented in the Quantum Espresso code.~\cite{PWSCF}
The generalized gradient approximation (GGA) was used for the exchange and correlation functional  
as parametrized by Perdew, Burke, and Ernzerhof.~\cite{PeBu96}
Static correlation effects were considered within the DFT$+U$ formalism~\cite{Anisimov:93}
using $U=4$ and $J=0.7$~eV for Ni~$3d$,
which is in line with previous work by us and others.~\cite{May:10, ABR:11, KimHan:15, Geisler-LNOSTO:17, WrobelGeisler:18}

In order to take octahedral tilts fully into account,
we have modeled the (LNO)$_n$/(LAO)$_n(001)$ SLs ($n=1,3$; denoted by 1/1 and 3/3 in the following)
by using $\sqrt{2}a \times \sqrt{2}a \times 2c$ $(6c)$ supercells,
rotated by $45^\circ$ around the \mbox{(pseudo-)}cubic $c$~axis,
that contain $20$ ($60$) atoms in total.
We have considered the effect of epitaxial strain
by setting the in-plane lattice parameter to
$a_\text{LAO} = 3.79~\AA$, $a_\text{STO} = 3.905~\AA$, or $a_\text{DSO} = 3.94~\AA$ (DyScO$_3$).
The out-of-plane lattice parameter has either been set
to previously determined values ($c = 3.93~\AA$ for $a_\text{LAO}$, $c = 3.83~\AA$ for $a_\text{STO}$)~\cite{ABR:11,Freeland:11}
or  has been optimized ($c = 3.79~\AA$ for $a_\text{DSO}$).
All structures exhibit an antiferrodistortive $a^-a^-c^-$ octahedral rotation pattern,
with a stronger effect in the NiO$_6$ octahedra (see Supplemental Material).
For $a_\text{STO}$ and $a_\text{DSO}$,
the AlO$_6$ octahedra show almost no rotation around the $c$~axis.

Wave functions and density have been expanded into plane waves up to cutoff energies of $35$ and $350$~Ry, respectively.
Ultrasoft pseudopotentials have been used, \cite{Vanderbilt:1990}
treating the
La $5s$, $5p$, $5d$, $6s$, $6p$,
Ni $3d$, $4s$,
Al $3s$, $3p$,
and O $2s$, $2p$
atomic subshells as valence states.
For La, Ni, and Al a non-linear core correction~\cite{LoFr82} has been included.
Different Monkhorst-Pack $\Vek{k}$-point grids~\cite{MoPa76} have been used
together with a Methfessel-Paxton smearing~\cite{MePa89} of $5$~mRy to sample the Brillouin zone:
$8 \times 8 \times 6$ for the 1/1 SLs and
$8 \times 8 \times 2$ for the 3/3 SLs.
The atomic positions have been optimized until the maximum component of the residual forces on the ions was less than $1$~mRy/a.u.
In the following, we explore the ferromagnetic phase of such SLs,
noting that also an antiferromagnetic state has been reported theoretically.~\cite{Puggioni:12}
A G-type antiferromagnetic order proved to be energetically less stable.

The electronic transport properties of the SLs have been calculated
from the DFT electronic structure on dense $\Vek{k}$-point grids
by using Boltzmann transport theory in the constant relaxation time approximation.
The BoltzTraP code~\cite{BoltzTraP:06} provides the energy- and spin-resolved transmission ${\cal T}_\sigma(E)$.
We obtained converged transmission curves by using
$64 \times 64 \times 24$ $\Vek{k}$ points for the 1/1 SLs and
$64 \times 64 \times 8$ $\Vek{k}$ points for the 3/3 SLs.
From the transmission we have calculated the thermoelectric quantities
by using the approach of Sivan and Imry,~\cite{SI86}
which is described in the Supplemental Material
and has been used in previous studies.~\cite{Geisler-LNOSTO:17, Geisler-Heusler:15, Gruner:15, GeislerPopescu:14, ComtesseGeisler:14, LamontagneGrunerPentcheva:16}

\section{\boldmath (LNO)$_1$/(LAO)$_1(001)$ superlattices}

\begin{figure}[tb]
	\centering
	\includegraphics{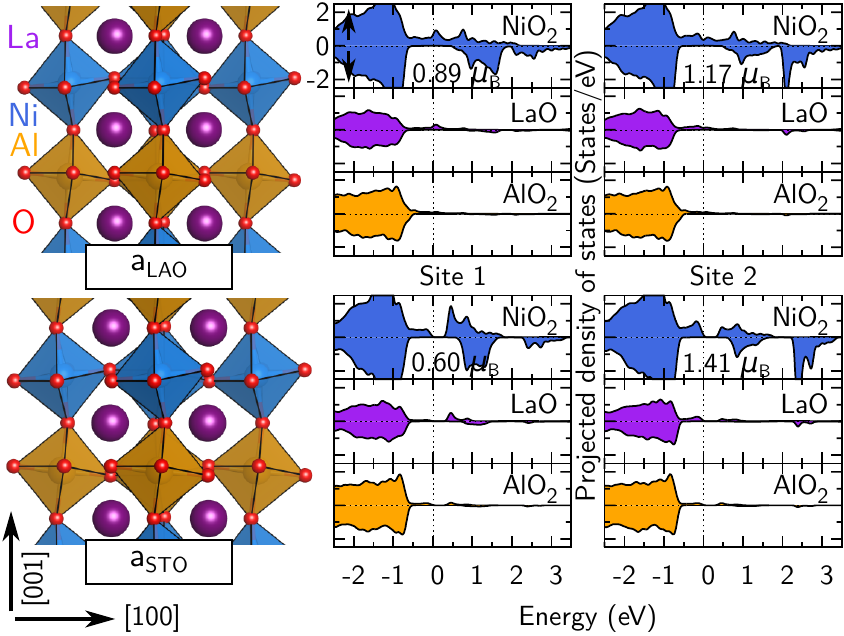}
	\caption{Side views of the optimized (LNO)$_1$/(LAO)$_1(001)$ SL structures and corresponding layer-, site-, and spin-resolved densities of states for compressive ($a_\text{LAO}$, top row) and tensile ($a_\text{STO}$, bottom row) epitaxial strain. The Ni spin magnetic moments are printed in the NiO$_2$ panels; their difference reflects the disproportionation. Zero energy denotes the Fermi energy ($a_{\text{LAO}}$) or the valence band maximum (VBM, $a_{\text{STO}}$). Note the band gap that opens in the case of tensile epitaxial strain.}
	\label{fig:SL-1-1-AtomicStructure}
\end{figure}

In bulk rare earth nickelates, charge disproportination (CD),
formally denoted as Ni$^{3+}$ $\to$ Ni$^{3+\delta}$ $+$ Ni$^{3-\delta}$,
has been discussed as the origin of the MIT.~\cite{Varignon:17, JohnstonCD:14, ParkMillisMarianetti:12, Mazin:07}
These effects also emerge in ultrathin films and heterostructures,~\cite{RENickelateReview:16} often in conjunction with structural breathing mode distortions (referred to as bond disproportionation).~\cite{ABR:11,LuHansmann:17}
In ultrathin 1/1 SLs,
such a MIT has been shown to be a consequence of the interplay of dimensional confinement through the wide-gap band insulating LAO spacer
and epitaxial strain.~\cite{ABR:11, Freeland:11}

\subsection{Nature of the metal-to-insulator transition and impact of octahedral tilts and disproportionation}

Figure~\ref{fig:SL-1-1-AtomicStructure} displays side views
of the optimized 1/1 SL structures
and corresponding layer-resolved densities of states (LDOS)
for two exemplary substrate lattice constants, $a_\text{LAO}$ (compressive strain) and $a_\text{STO}$ (tensile strain).
The LDOS plots show that the states around the Fermi level are localized in the NiO$_2$ layers.
Moreover, while the system at $a_\text{LAO}$ is metallic, a band gap of $0.29$~eV opens for tensile strain.
Consistent with previous results,~\cite{ABR:11} this is caused by a disproportionation at the Ni sites,
expressed in the variation of the Ni1/Ni2 magnetic moments,
$0.89/1.17~\mu_\text{B}$ ($a_\text{LAO}$) and $0.60/1.41~\mu_\text{B}$ ($a_\text{STO}$),
and accompanied by distinct Ni-O bond lengths. (See the Supplemental Material for further structural information.)

\begin{figure}[tb]
	\centering
	\includegraphics{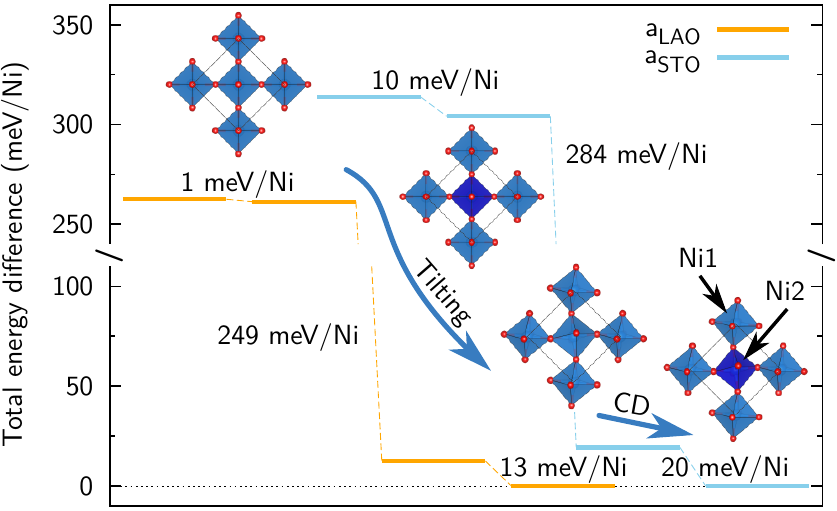}
	\caption{Stability diagram of (LNO)$_1$/(LAO)$_1(001)$ SLs for different structural and electronic phases and varying epitaxial strain. The respective ground state energies have been chosen as references. The sequence is, from left to right: undistorted system; CD without tilts; tilted without CD; tilted with CD. Different octahedral colors (light/dark blue) are used here to highlight the Ni1/Ni2 disproportionation within the nickelate $(001)$ planes.}
	\label{fig:SL-1-1-EnergyDifferences}
\end{figure}

To disentangle the impact of octahedral tilting and CD,
Fig.~\ref{fig:SL-1-1-EnergyDifferences} compares total energies
between different structural phases of the 1/1 SLs, starting from an undistorted configuration
and comparing the separate and total effect of octahedral tilting and CD under compressive and tensile strain. 
The energy differences are generally higher
for $a_\text{STO}$ than for $a_\text{LAO}$.
The largest energy gain
(about $250$/$280$~meV/Ni for $a_\text{LAO}$/$a_\text{STO}$)
is caused by octahedral tilting.
Disproportionation lowers the total energy
on a smaller scale
and stronger in the case of a tilted structure
($13$/$20$~meV/Ni for $a_\text{LAO}$/$a_\text{STO}$)
than in the untilted case
($1$/$10$~meV/Ni, respectively).
Hence, octahedral tilts increase the tendency towards disproportionation,
and the effect is stronger for tensile than for compressive epitaxial strain.
We note that the inequivalence of the Ni sites is mandatory
for the opening of a band gap.

\begin{figure}[!tb]
	\centering
	\includegraphics{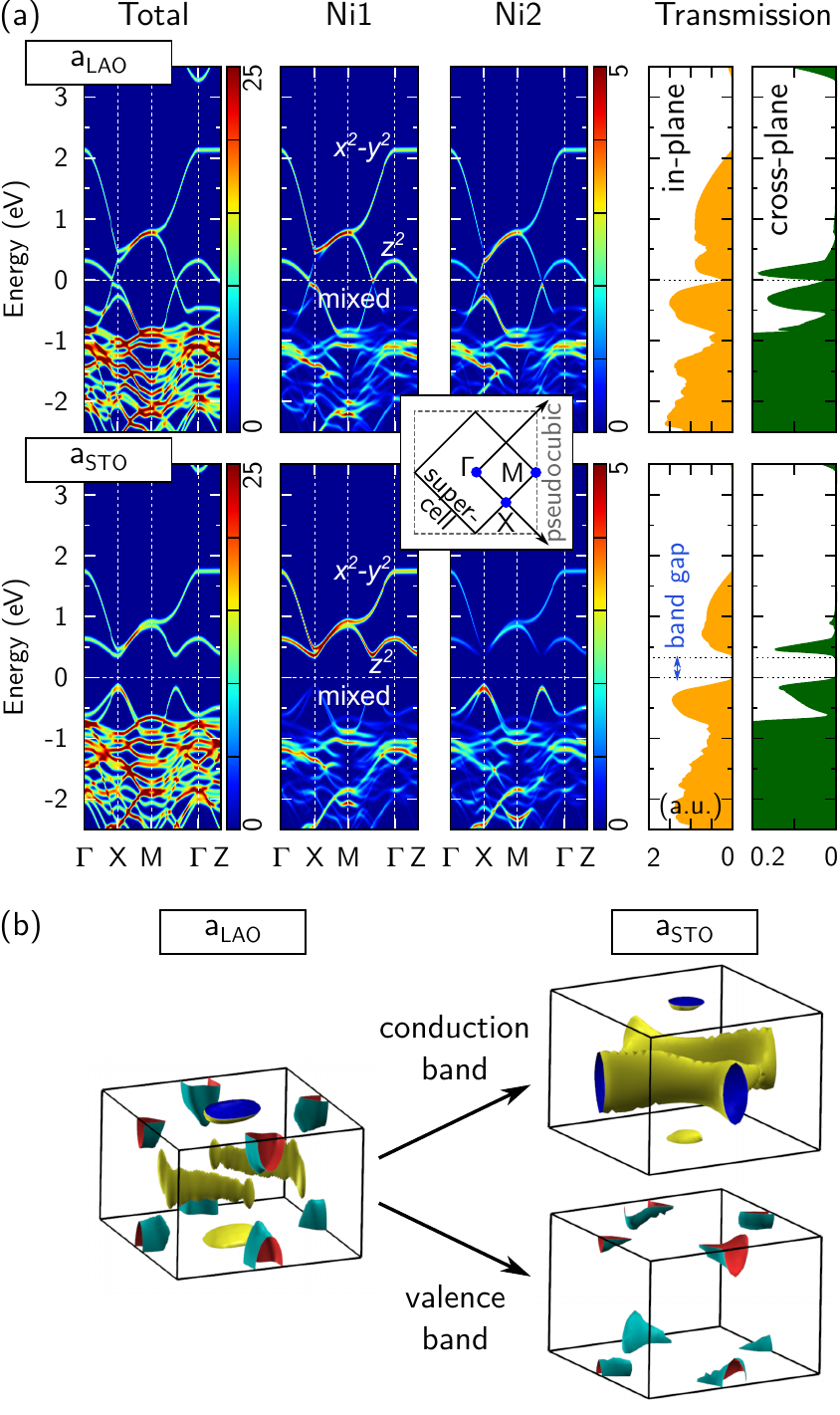}
	\caption{(a)~Majority spin band structures, projections onto the two different Ni sites, respective $3d$ orbital characters, and corresponding transmission~${\cal T}_\uparrow(E)$ plots of (LNO)$_1$/(LAO)$_1(001)$ SLs for compressive ($a_\text{LAO}$) and tensile ($a_\text{STO}$) epitaxial strain. Zero energy denotes the Fermi energy ($a_{\text{LAO}}$) or the VBM ($a_{\text{STO}}$). The inset clarifies the relation between the \mbox{(pseudo-)}cubic and our supercell Brillouin zone.
	(b)~Fermi-surface-like sheets taken in the vicinity of the Fermi energy ($a_{\text{LAO}}$) or near the band edges ($a_{\text{STO}}$) illustrate the overlap of bands being lifted due to increasing epitaxial strain.}
	\label{fig:SL-1-1-Bands}
\end{figure}

Figure~\ref{fig:SL-1-1-Bands}(a) analyzes further how the vertical confinement
splits the four Ni $3d$ $e_g$ majority-spin bands
(which are degenerate in metallic bulk LNO)
into two sets around the Fermi energy $E_\text{F}$.
While the set below $E_\text{F}$ has mixed $d_{z^2}$ and $d_{x^2-y^2}$ character, the set above the Fermi level comprises a lower lying narrower band of $d_{z^2}$ character, followed by a more dispersive $d_{x^2-y^2}$ band at higher energies. In the case of compressive strain, this distinction is less strict,
and the sets of bands touch or even overlap slightly, for instance close to the $X$ or $Z$ point or between $\Gamma$ and $M$. Under tensile strain, however, these two sets of bands split, opening a gap of $0.29$~eV at $a_\text{STO}$.
In this case, the unoccupied states up to $2$~eV are derived predominantly from the Ni1 ions,
whereas the occupied states between $E_\text{F}$ and $E_\text{F}-0.6$~eV stem mostly from the Ni2 ions.
The contribution of both sets to the Fermi surface for $a_\text{LAO}$ is clearly visible in Fig.~\ref{fig:SL-1-1-Bands}(b) and splits to valence band maximum (VBM) and conduction band minimum (CBM) for tensile strain. 
Such complex electronic energy isosurfaces exhibiting many pockets are typical for good thermoelectric materials.~\cite{Xing:16}

Both sets of bands retain a considerable degree of dispersion,
which leads to a good electronic transmission~${\cal T}_\uparrow(E)$
in-plane ($xy$) and cross-plane ($z$),
depending on the type of Ni~$3d$ orbitals involved [Fig.~\ref{fig:SL-1-1-Bands}(a)].
Note the two steep transmission peaks (or hills)
near the Fermi energy.
In-plane, the peak below $E_\text{F}$
shows a stronger, more concentrated transmission than the peak above $E_\text{F}$.
Both peaks stem from the Ni~$3d_{x^2-y^2}$ orbitals.
For cross-plane transport, the peaks are similar in height,
but the peak above $E_\text{F}$ is narrower.
In this case, they are related to the Ni~$3d_{z^2}$ orbitals.
Due to the almost insulating AlO$_2$ layer (cf.~Fig.~\ref{fig:SL-1-1-AtomicStructure}),
the cross-plane transmission is always smaller
than the in-plane transmission.

\subsection{Thermoelectric properties}

\begin{figure}[tbh!]
	\centering
	\includegraphics{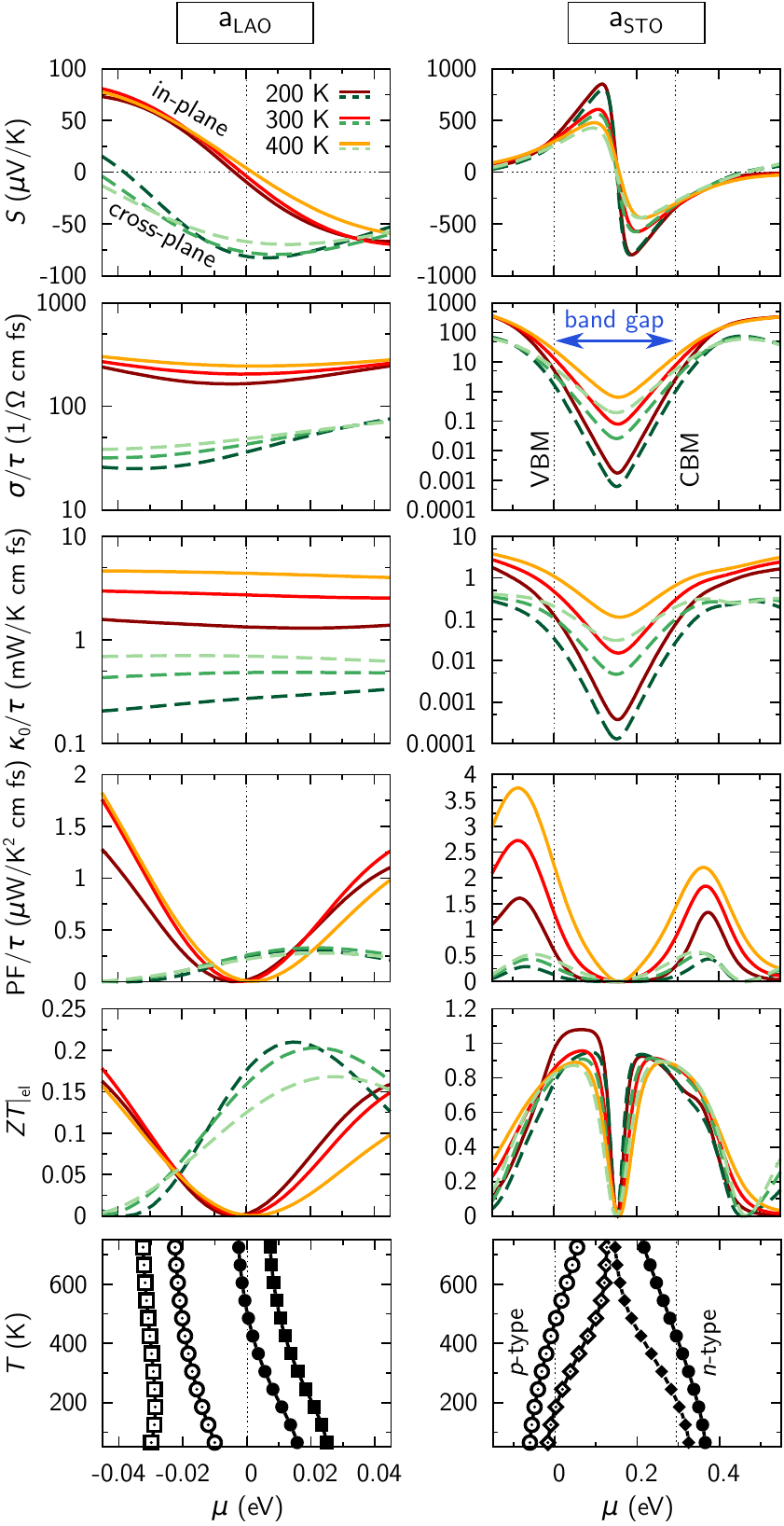}
	\caption{Thermoelectric properties of (LNO)$_1$/(LAO)$_1(001)$ SLs for compressive ($a_\text{LAO}$) and tensile ($a_\text{STO}$) epitaxial strain at three different temperatures. From top to bottom, Seebeck coefficient~$S$,
electrical conductivity $\sigma/\tau$,
electronic contribution to the thermal conductivity $\kappa_0/\tau$,
power factor PF$/\tau$,
and electronic figure of merit~$ZT|_\text{el}$ are shown,
where $\tau$ denotes the relaxation time constant. Red-orange solid (green dashed) lines depict in-plane (cross-plane) transport. Zero energy denotes the Fermi energy ($a_{\text{LAO}}$) or the VBM ($a_{\text{STO}}$). In the latter case, the CBM is also marked. For the same energy range, the variation of the chemical potential $\mu(T)$ with the temperature is shown for six different exemplary doping levels: $\pm 1 e^-$ (diamonds), $\pm 10 e^-$ (circles), and $\pm 20 e^-$ (squares) per $10^4$ atoms.}
	\label{fig:SL-1-1-TE}
\end{figure}

The thermoelectric properties of the 1/1 SLs are summarized in Fig.~\ref{fig:SL-1-1-TE}
for compressive ($a_\text{LAO}$) and tensile ($a_\text{STO}$) epitaxial strain.
(Further results for $a_\text{DSO}$ are shown in the Supplemental Material).
The layered crystal structure would imply  
an anisotropic thermoelectric response in the 1/1 SLs,
i.e., different in- and cross-plane properties,
as for example a Seebeck coefficient of $+135$~$\mu$V/K cross-plane and $-15$~$\mu$V/K in-plane predicted in (LNO)$_3$/(STO)$_3(001)$ SLs with
$p$-type interfaces at room temperature,~\cite{Geisler-LNOSTO:17}
or $\sim-300$~$\mu$V/K cross-plane,
but only $\sim+10$~$\mu$V/K in-plane for the layered delafossite PtCoO$_2$ under tensile strain.~\cite{Gruner:15}
Surprisingly, we obtain
large and very similar Seebeck coefficients \textit{both} in-plane and cross-plane,
in particular for $a_\text{STO}$
with values reaching $\sim \pm 825$, $\pm 600$, and $\pm 450$~$\mu$V/K in- and cross-plane
at $200$, $300$, and $400$~K, respectively (Fig.~\ref{fig:SL-1-1-TE}).
The unusually high values are traced back to the confinement-driven MIT,
i.e., the opening of a band gap
due to the coupling of short-period vertical SL design and tensile epitaxial strain.
The band gap is accompanied by highly asymmetric
in- and cross-plane transmissions~${\cal T}(E)$
near the chemical potential~$\mu$,
which result in large in- and cross-plane Seebeck coefficients.
Hence, it is a remarkable feature of the 1/1 SLs
that their in-plane thermoelectric response is strongly controlled
by quantum confinement
in the perpendicular $[001]$ direction.
This aspect will be further corroborated when we discuss later the 3/3 SLs,
which exhibit a different degree of confinement.

The bands around the Fermi energy constitute a good mixture
between being localized (high asymmetry/Seebeck coefficient, but low conductivity)
and delocalized (high conductivity, but low asymmetry/Seebeck coefficient).
For tensile strain, the opening of the band gap
is accompanied by an increased steepness
of the transmission peaks at its edges [Fig.~\ref{fig:SL-1-1-Bands}(a)],
which further enhances the Seebeck coefficient.
In contrast,
compensation of $n$- and $p$-type contributions due to the closed band gap
reduces the Seebeck coefficient for compressive strain to $\sim \pm 80$~$\mu$V/K.
For $a_\text{LAO}$, the three Seebeck curves for different temperatures
are rather close,
whereas for $a_\text{STO}$ they become flatter
(lower Seebeck coefficients attainable)
with increasing temperature (Fig.~\ref{fig:SL-1-1-TE}).

The electrical conductivities
($\sigma \approx 10^{-3}$ to $10^{3} / \Omega \, \text{cm}$
assuming $\tau$ to be in the fs range, Fig.~\ref{fig:SL-1-1-TE})
are in the typical range of doped semiconductors
and increase with temperature.
Owing to the transmission anisotropy [Fig.~\ref{fig:SL-1-1-Bands}(a)],
they are almost one order of magnitude smaller
for cross-plane transport than for in-plane transport.
While $\sigma(\mu)$ is nearly constant in the shown energy interval
for compressive strain,
it exhibits a drastic reduction within the emerging band gap
for tensile strain.
The electronic contribution to the thermal conductivity~$\kappa_0$
is found to always follow the trend of the electrical conductivity~$\sigma$
(Fig.~\ref{fig:SL-1-1-TE}).

The higher Seebeck coefficients arising for $a_\text{STO}$
relative to $a_\text{LAO}$
result in higher power factors $\text{PF} = \sigma S^2$.
(We note that in Fig.~\ref{fig:SL-1-1-TE} we report PF/$\tau$ to avoid the ambiguity of choice of the relaxation time).
For $a_\text{STO}$, PF/$\tau$ increases considerably with temperature,
in particular for in-plane transport
(e.g., $1.7$, $2.75$, and $3.75$~$\mu$W/K$^2$cm\,fs near the VBM for $200$, $300$, and $400$~K, respectively).
This is due to the strong increase of~$\sigma$ with temperature,
which overcompensates the decrease of~$S^2$.
The larger in-plane power factors near the VBM relative to those near the CBM
are related to the different width and height of the corresponding transmission peaks, as mentioned above
[Fig.~\ref{fig:SL-1-1-Bands}(a)].
Due to the anisotropic conductivities,
the power factors are mostly higher in-plane than cross-plane
(where they reach values around $0.5$~$\mu$W/K$^2$cm\,fs).

Further increase of tensile strain
(i.e., by using DSO instead of STO as substrate)
improves the Seebeck coefficient within the band gap,
but not necessarily near the band edges (see Supplemental Material).
Moreover, it lowers the power factor for both the in- and the cross-plane case.
Hence, the thermoelectric response is better on STO,
particularly in-plane.
We expect further performance improvement
on substrates with lattice parameters
slightly \textit{lower} than that of STO,
like LaGaO$_3$ or NdGaO$_3$.
An additional advantage of such substrates
is the nonpolar interface to the SL.

The electronic figure of merit $ZT|_\text{el} = \sigma S^2 T / \kappa_{0}$
reaches significantly larger values
for tensile strain ($0.8$ -- $1.1$) than for compressive strain ($0.2$), a trend observed also for the power factor
(Fig.~\ref{fig:SL-1-1-TE}).
Moreover, for $a_\text{STO}$, the in- and cross-plane $ZT|_\text{el}$ values are similar.
This stems from the peculiar quasi-equivalence of the in- and cross-plane Seebeck coefficients
and the similar trend of $\sigma$ and $\kappa_0$.

The position of the chemical potential plays an important role
for the thermoelectric properties.~\cite{Geisler-Heusler:15}
To provide an idea of its variability,
we have calculated $\mu(T)$ 
from the rigid electronic structure (Fig.~\ref{fig:SL-1-1-TE}),
using six different exemplary doping levels:
$\pm 1$, $\pm 10$, and $\pm 20 e^-$ per $10^4$ atoms,
corresponding to $\sim \pm 8.6 \cdot 10^{18}$ -- $\pm 1.7 \cdot 10^{20}$ $e^{-}$/cm$^{3}$.
Particularly for the case of $a_\text{STO}$,
where a finite band gap exists,
the chemical potential can be varied
over a wide energy range
(e.g., by extrinsic $n$- or $p$-type doping),
such as to
optimize the thermoelectric properties
in the temperature interval of interest.

\begin{table}[tb]
	\centering
	\vspace{-1.5ex}
	\caption{\label{tab:SL-1-1-TE-Comparison}Comparison of the attainable thermoelectric performance of different oxide systems around $T=300$~K. In the case of theoretical results, the used relaxation times~$\tau$ are given. Note that Seebeck coefficient~$S$ and power factor~PF usually reach their maxima under different conditions, e.g., different doping levels (see Supplemental Material).}
	\begin{ruledtabular}
	\begin{tabular}{lcc}
		System & $S$ ($\mu$V/K) & PF ($\mu$W/K$^2$cm)	\\
		\hline
		\multicolumn{3}{l}{(LNO)$_1$/(LAO)$_1$ (this work, $\tau = 4$~fs)}	\\
		$a_\text{LAO}$, in-plane		& $-70$, $+80$		& $7.2$	\\
		$a_\text{STO}$, in-plane		& $\pm 600$			& $11.0$	\\
		$a_\text{DSO}$, in-plane		& $\pm 740$			& $8.0$	\\
		$a_\text{LAO}$, cross-plane		& $-80$				& $1.2$	\\
		$a_\text{STO}$, cross-plane		& $\pm 600$			& $2.0$	\\
		$a_\text{DSO}$, cross-plane		& $\pm 740$			& $1.6$	\\
		\hline
		LNO (this work, $\tau = 4$~fs)										& $-12$				& $1.6$	\\
		LNO (exp.~\cite{LNO-LCO-Thermo:14})									& $-18$				& $3.2$	\\
		\hline
		STO (DFT,~\cite{GhosezSTO:16} $\tau = 4.3$~fs)						& $-400$			& $10$	\\
		La:STO bulk (exp.~\cite{Okuda:01})									& $-380$			& $35$	\\
		La:STO films (exp.~\cite{Stemmer:10})								& $-980$			& $39$	\\
		\hline
		CCO (DFT,~\cite{GhosezCCO:17} $\tau = 0.8$~fs)									& $-500$, $+700$	& $0.7$	\\
		CCO @ $a_\text{STO}$ (DFT,~\cite{CCO-Schwingen-APL:17} $\tau = 0.3$~fs)			& $+25$				& $0.3$	\\
		CCO/STO (exp.~\cite{YordanovCCO:17})											& $+150$			& $3$	\\
	\end{tabular}
	\end{ruledtabular}
\end{table}

In Table~\ref{tab:SL-1-1-TE-Comparison} we compare the 1/1 SLs to some of the best performing oxide systems reported so far.
(More details can be found in the Supplemental Material.)
To evaluate the power factor, we have derived the relaxation time by comparing the DFT$+U$ conductivity at $T=300$~K for paramagnetic bulk LNO,
$\sigma / \tau \approx 2500 / \Omega \, \text{cm} \, \text{fs}$,
to the experimental value~\cite{LNO-LCO-Thermo:14} $\sigma_\text{exp} \approx 10^4 / \Omega \, \text{cm}$.
The obtained $\tau=4$~fs is within the typical range of relaxation times;
for instance, it is close to the value used for bulk STO (Table~\ref{tab:SL-1-1-TE-Comparison}).~\cite{GhosezSTO:16}
We will later provide further evidence that supports this value.
Note that we expect the real~$\tau$ to be significantly anisotropic for the SLs
due to their layered crystal structure.
For instance, values of $70$~fs (in-plane) and $12$~fs (cross-plane)
have been reported for PtCoO$_2$ at room temperature.~\cite{Gruner:15}
In any case, the Seebeck coefficient~$S$
and the electronic figure of merit~$ZT|_\text{el}$ are independent of $\tau$.

For $a_\text{STO}$ and $a_\text{DSO}$,
both the in- and cross-plane Seebeck coefficients
of the 1/1 SLs clearly exceed those
of bulk STO obtained from DFT calculations~\cite{GhosezSTO:16}
or those measured for La-doped bulk STO samples,~\cite{Okuda:01}
but are lower than the values measured for 
La-doped STO films~\cite{Stemmer:10} (Table~\ref{tab:SL-1-1-TE-Comparison}).
Similarly, the 1/1 SLs exhibit higher power factors than bulk STO; however,
the La-doped STO systems can reach values three times as high.

Another promising materials class for thermoelectric applications is cobaltates
such as the mentioned Pd/Pt-based delafossites~\cite{Gruner:15}
or, in particular, the misfit-layered oxide Ca$_3$Co$_4$O$_9$ (CCO).~\cite{YordanovCCO:17, GhosezCCO:17, CCO-Schwingen-APL:17}
Remarkably, the 1/1 SLs can compete with this material,
regardless of the underlying substrate (Table~\ref{tab:SL-1-1-TE-Comparison}).
For $a_\text{STO}$ and $a_\text{DSO}$,
both the in- and cross-plane Seebeck coefficients
of the 1/1 SLs are close to (or even exceed) the in-plane DFT results for pristine CCO,
which are already much higher than DFT results for CCO strained to $a_\text{STO}$.
The in-plane power factors of the 1/1 SLs
are by far better than those obtained for the different CCO systems,
also for compressive strain ($a_\text{LAO}$).
Even the cross-plane power factors of the 1/1 SLs
reach values larger than or close to the in-plane results
for the different CCO systems,
whose cross-plane thermoelectric response is negligible.

Note that the DFT power factors listed in Table~\ref{tab:SL-1-1-TE-Comparison}
tend to be lower than experimentally measured ones. Thus,
we expect an even better thermoelectric performance for grown 1/1 SLs.

\section{\boldmath (LNO)$_3$/(LAO)$_3(001)$ superlattices}

In order to demonstrate the crucial effect of confinement on the thermoelectric properties,
we now increase the thickness of both the LNO and the LAO spacer layer to $n=3$.

\begin{figure}[tb]
	\centering
	\includegraphics{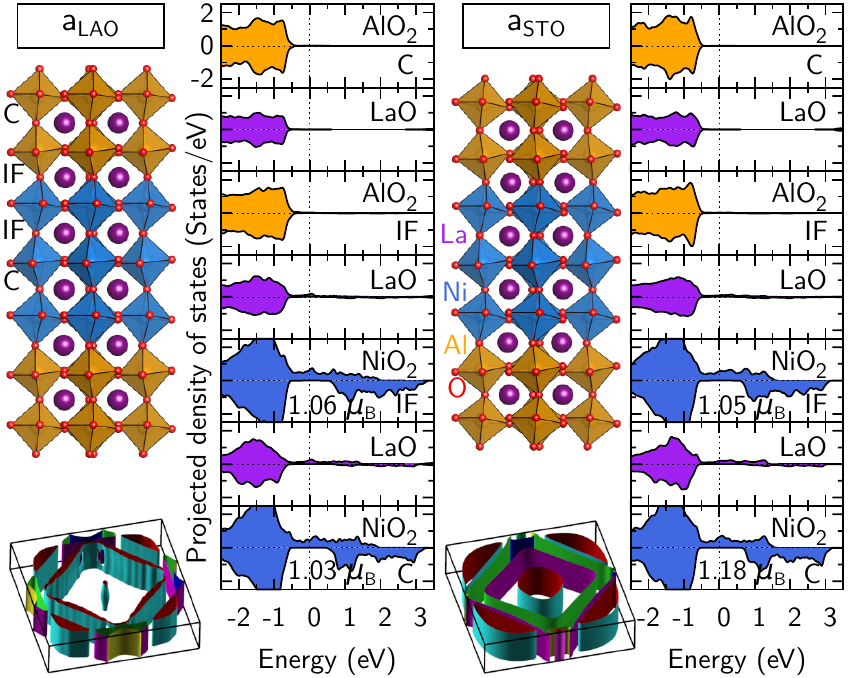}
	\caption{Optimized structures and layer- and spin-resolved densities of states of (LNO)$_3$/(LAO)$_3(001)$ SLs for compressive ($a_\text{LAO}$, left) and tensile ($a_\text{STO}$, right) epitaxial strain. (Colors and axes as in Fig.~\ref{fig:SL-1-1-AtomicStructure}. C~and IF denote central and interfacial layers, respectively.) The Ni spin magnetic moments are printed in the NiO$_2$ panels. Zero energy denotes the Fermi energy. In addition, the Fermi surfaces are shown.}
	\label{fig:SL-3-3-AtomicStructure}
\end{figure}

\begin{figure}[tbh!]
	\centering
	\includegraphics{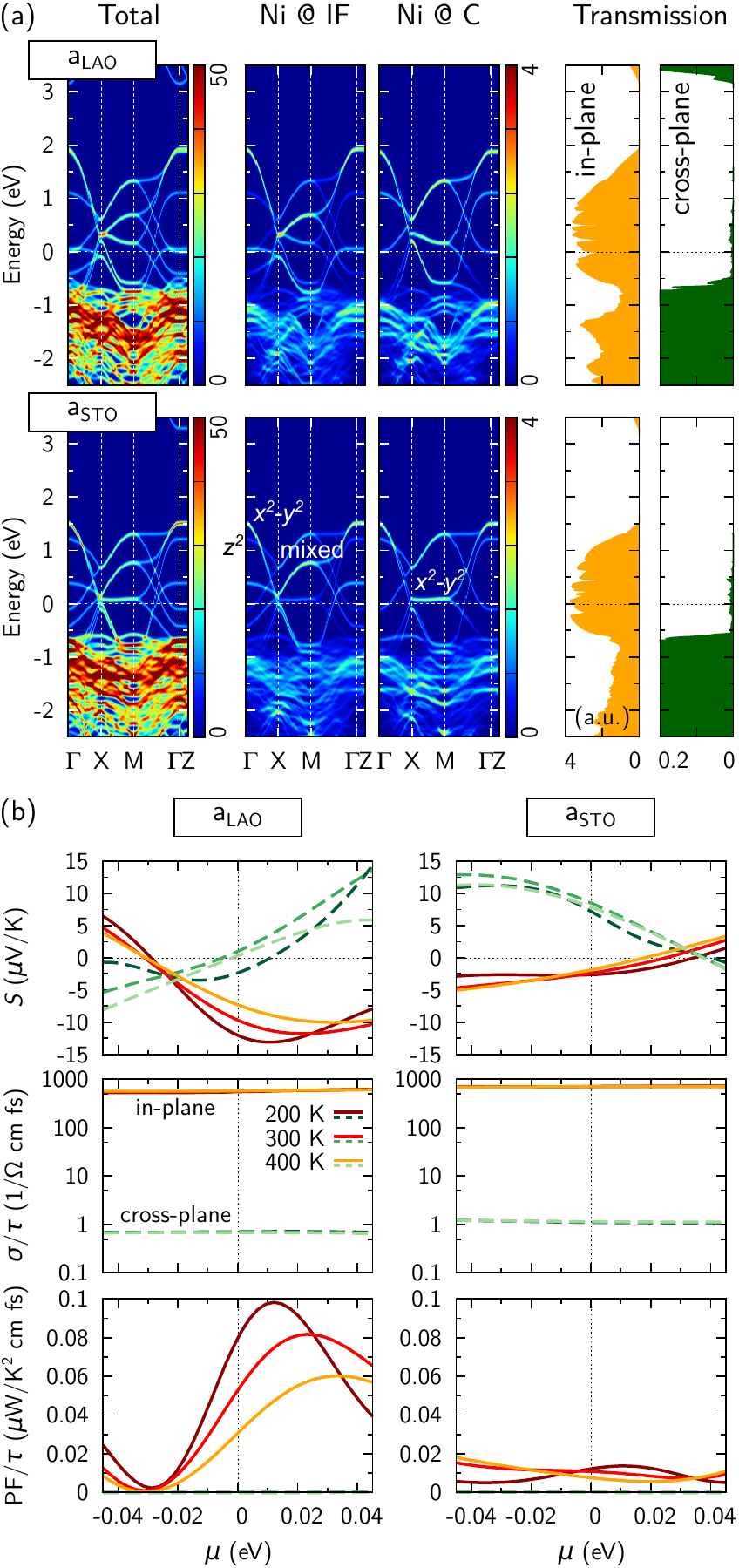}
	\caption{(a)~Majority spin band structures, projections onto the interfacial and central Ni sites (cf.~Fig.~\ref{fig:SL-3-3-AtomicStructure}), respective $3d$ orbital characters, and corresponding transmission~${\cal T}_\uparrow(E)$ plots of (LNO)$_3$/(LAO)$_3(001)$ SLs for compressive ($a_\text{LAO}$) and tensile ($a_\text{STO}$) epitaxial strain. (b)~Corresponding thermoelectric properties at three different temperatures. Red-orange solid (green dashed) lines depict in-plane (cross-plane) transport.}
	\label{fig:SL-3-3-BandsTE}
\end{figure}

\subsection{Electronic structure}

The optimized structures of the 3/3 SL for compressive and tensile strain are displayed in
Fig.~\ref{fig:SL-3-3-AtomicStructure}.
In agreement with previous studies,~\cite{ABR:11} the LDOS plots show that the LNO layer is now metallic,
irrespective of the degree of epitaxial strain and with no tendency towards disproportionation at the Ni sites. 
Manually disproportionated in-plane Ni-O bonds relaxed back to equal Ni-O bonds.
(Structural information can be found in the Supplemental Material.)

Similarly to (LNO)$_3$/(STO)$_3(001)$ SLs,~\cite{Geisler-LNOSTO:17}
we observe the formation of
Ni-$3d_{x^2-y^2}$- and Ni-$3d_{z^2}$-derived quantum well states
within the LAO band gap
that respond differently to epitaxial strain;
i.e., increasing the lattice parameter~$a$ from $a_\text{LAO}$ to $a_\text{STO}$
significantly lowers the energies of the former states
while slightly raising that of the latter states [Fig.~\ref{fig:SL-3-3-BandsTE}(a)].
No significant dispersion arises in the cross-plane direction
owing to the strong quantum confinement by the thick LAO barrier,
which has an even larger band gap than STO.
The Fermi surfaces of the 3/3 SLs shown in Fig.~\ref{fig:SL-3-3-AtomicStructure}
exhibit a quasi-two-dimensional, cylindrical shape,
in contrast to those of the 1/1 SLs (cf.~Fig.~\ref{fig:SL-1-1-Bands}).
Due to the absence of charge doping,
the Fermi surface of the 3/3 SL for $a_\text{STO}$
strongly resembles that of a (LNO)$_3$/(STO)$_3(001)$ SL for $a_\text{STO}$
and $np$-type interface coupling.~\cite{Geisler-LNOSTO:17}

The response of the Ni magnetic moments to epitaxial strain
is much weaker for the 3/3 SLs than for the 1/1 SLs
(Fig.~\ref{fig:SL-3-3-AtomicStructure}).
Generally, they are close to the LNO bulk value ($1.04~\mu_\text{B}$),
the only significant increase being found for the central Ni ion
in the case of tensile strain ($1.18~\mu_\text{B}$).
Since no disproportionation arises in this system,
the Ni sites within each plane are equivalent.

\subsection{Thermoelectric properties}

\begin{figure}[tb]
	\centering
	\includegraphics{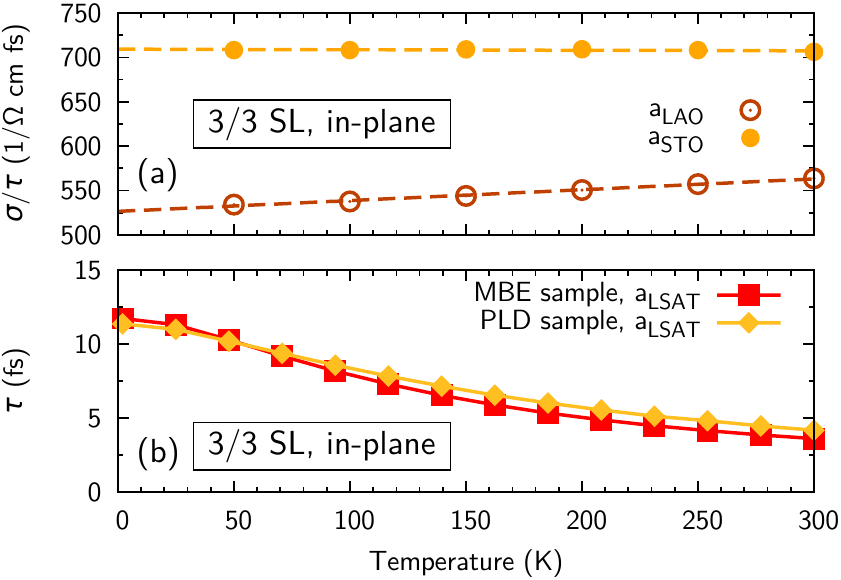}
	\caption{(a)~Simulated in-plane electrical conductivity of (LNO)$_3$/(LAO)$_3(001)$ SLs for compressive ($a_\text{LAO}$) and tensile ($a_\text{STO}$) epitaxial strain [cf.~Fig.~\ref{fig:SL-3-3-BandsTE}(b), $\mu = E_\text{F}$]. (b)~Estimated in-plane relaxation times obtained by relating simulated and measured~\cite{Wrobel-LNOLAO:17} electrical conductivities.}
	\label{fig:SL-3-3-SigmaTau}
\end{figure}

The two-dimensional metallic nature of the 3/3 SLs
leads to a high in-plane electrical conductivity [Figs.~\ref{fig:SL-3-3-BandsTE}(b) and~\ref{fig:SL-3-3-SigmaTau}(a)],
which exceeds that of the 1/1 SLs (cf.~Fig.~\ref{fig:SL-1-1-TE})
and is higher for $a_\text{STO}$ ($\sim 710 / \Omega \, \text{cm} \, \text{fs}$)
than for $a_\text{LAO}$ ($\sim 550 / \Omega \, \text{cm} \, \text{fs}$).
It shows only a relatively weak dependence on temperature and chemical potential.

Wrobel \textit{et al.}\ recently measured the in-plane electrical conductivity
of 3/3 SLs grown by molecular beam epitaxy (MBE) and pulsed laser deposition (PLD)
on (LaAlO$_3$)$_{0.3}$(Sr$_2$AlTaO$_6$)$_{0.7}$ (LSAT, $a_\text{LSAT} = 3.87~\AA$).~\cite{Wrobel-LNOLAO:17}
By comparing their measurements to our simulated results
[Figs.~\ref{fig:SL-3-3-BandsTE}(b) and~\ref{fig:SL-3-3-SigmaTau}(a),
accounting for the different substrate lattice parameters via averaging],
we obtained estimated temperature-dependent relaxation times $\tau(T)$ as shown in Fig.~\ref{fig:SL-3-3-SigmaTau}(b).
The curves decay monotonically with temperature,
most likely due to increasing electron-phonon scattering.
MBE- and PLD-grown samples give quite similar relaxation times,
ranging from $4$ to $12$~fs in the present temperature interval.
These results provide further confidence
that $\tau = 4$~fs estimated above from bulk LNO at $300$~K
is a reasonable approximation for the present SLs.

The thermoelectric response of the 3/3 SLs is much smaller than that of the 1/1 SLs,
which is caused by their metallic character,
i.e, the highly symmetric transmission around the Fermi energy.
In- and cross-plane Seebeck coefficients are of comparable magnitude
and do not exceed $15$~$\mu$V/K in absolute value.
Consequently,
the highest power factor at room temperature,
$0.32$~$\mu$W/K$^2$cm
[obtained in-plane for $a_\text{LAO}$ with $\tau \approx 4$~fs from Fig.~\ref{fig:SL-3-3-SigmaTau}(b)],
is distinctly below those of the 1/1 SLs listed in Table~\ref{tab:SL-1-1-TE-Comparison}.
At $200$~K, it maximally reaches $0.6$~$\mu$W/K$^2$cm ($\tau \approx 6$~fs).
The cross-plane conductivity is considerably lower than for the 1/1 SLs
due to the thicker insulating LAO barrier,
which leads to even smaller power factors cross-plane than in-plane.

\vspace{-2ex}

\section{Summary}

We investigated the thermoelectric response
of (LaNiO$_3$)$_n$/(LaAlO$_3$)$_n(001)$ superlattices ($n=1,3$)
by combining density functional theory calculations
including an on-site Coulomb repulsion term
and Boltzmann theory in the constant relaxation time approximation.
The thermoelectricity of the $n=3$ system was found to be impeded
by the two-dimensional metallic nature of the LaNiO$_3$ region
and the too thick insulating LaAlO$_3$ spacer layer.
In contrast,
the reduction of both layers to the ultrathin $n=1$ limit
strongly enhances the thermoelectric properties of the superlattice,
particularly in-plane.
This originates from the confinement-induced Ni-site disproportionation,
which is accompanied by a band splitting (gap opening) at the Fermi energy
and is promoted considerably
by tensile epitaxial strain and octahedral tilting,
as shown from disentangling the impact
of these three aspects on the total energy.
Due to the sensitivity of the metal-to-insulator transition
(i.e., the width of the band gap)
and of the band velocities near the Fermi energy
to epitaxial strain,
the latter emerges as a control parameter
that allows us to optimize the thermoelectric performance.
For superlattices under tensile strain
corresponding to the SrTiO$_3$ lattice parameter,
we predicted in- and cross-plane Seebeck coefficients of $\pm 600$~$\mu$V/K
and an in-plane power factor of $11$~$\mu$W/K$^2$cm
using an estimated relaxation time of $\tau = 4$~fs
around room temperature;
the power factor showed even further increase with temperature.
These values compare well to prominent oxide thermoelectrics
like La-doped SrTiO$_3$ or layered cobaltates such as Ca$_3$Co$_4$O$_9$
and can be traced back to the opening of a small band gap ($0.29$~eV).

Although the growth of oxide superlattices with atomic precision is more challenging than for (doped) bulk oxides,
it has become a well-established procedure
over the past decade~\cite{Freeland:11, Boris:11, Benckiser:11, Wrobel-LNOLAO:17, Middey:18}
and its application for devices is intensively pursued.
Our findings suggest that oxide heterostructures at the verge of a metal-to-insulator transition
are promising candidates for thermoelectric materials.
This opens new avenues for research on designed superlattices for thermoelectric applications.

\vspace{-2ex}

\begin{acknowledgments}

We thank Eva Benckiser (Stuttgart) for sharing the
(LaNiO$_3$)$_3$/(LaAlO$_3$)$_3(001)$ electronic transport results.
This work was supported by the German Science Foundation (Deutsche Forschungsgemeinschaft, DFG) within the SFB/TRR~80, Projects No.\ G3 and No.\ G8.
Computing time was granted by the Center for Computational Sciences and Simulation of the University of Duisburg-Essen
(DFG Grants No.\ INST 20876/209-1 FUGG and No.\ INST 20876/243-1 FUGG).

\end{acknowledgments}


\end{document}